\documentclass[twocolumn,amsmath,amssymb]{revtex4}

\usepackage{natbib}
\usepackage{epsfig}
\usepackage{amsmath,amssymb,mathrsfs}

\newcommand{\rrr}{{\bf{r}}}
\newcommand{\rrrp}{{\bf{r}}_\perp}
\newcommand{\nablabf}{\boldsymbol{\nabla}}
\newcommand{\ex}{{{\bf e}}_x}
\newcommand{\ey}{{{\bf e}}_y}
\newcommand{\ez}{{{\bf e}}_z}

\newcommand{\vvv}{{{\bf v}^{{}}}}
\newcommand{\EEE}{{{\bf E}^{{}}}}

\begin{document}

\title{Mass and charge transport in micro and nano-fluidic
channels\footnote{Invited paper presented at the Second
International Conference on Transport Phenomena in Micro and
Nanodevices, Il Ciocco Hotel and Conference Center, Barga, Italy,
11-15 June 2006. Accepted for publication in a special issue of
\emph{Nanoscale and Microscale Thermophysical Engineering} (Taylor
\& Francis).}}

\author{Niels Asger Mortensen,\footnote{Corresponding author. Email: nam@mic.dtu.dk, URL: www.mic.dtu.dk/nam, Phone: +45 4525 5724, Fax: +45 4588 7762} Laurits H. Olesen, Fridolin Okkels, and Henrik Bruus}

\affiliation{MIC -- Department of Micro and Nanotechnology, NanoDTU,\\
Technical University of Denmark, DK-2800 Kongens Lyngby, Denmark }

\date{\today}

\begin{abstract}
We consider laminar flow of incompressible electrolytes in long,
straight channels driven by pressure and electro-osmosis. We use a
Hilbert space eigenfunction expansion to address the general
problem of an arbitrary cross section and obtain general results
in linear-response theory for the mass and charge transport
coefficients which satisfy Onsager relations. In the limit of
non-overlapping Debye layers the transport coefficients are simply
expressed in terms of parameters of the electrolyte as well as the
hydraulic radius ${\cal R}=2{\cal A}/{\cal P}$ with $\cal A$ and
$\cal P$ being the cross-sectional area and perimeter,
respectively. In particular, we consider the limits of thin
non-overlapping as well as strongly overlapping Debye layers,
respectively, and calculate the corrections to the hydraulic
resistance due to electro-hydrodynamic interactions.
\end{abstract}


 \maketitle

\section{Introduction}

Laminar Hagen--Poiseuille and electro-osmotic flows are important
to microfluidics and a variety of lab-on-a-chip
applications~\cite{Laser:04,Stone:04a,Squires:05a} and the rapid
development of micro and nano fabrication techniques is putting
even more emphasis on flow in channels with a variety of shapes
depending on the fabrication technique in use. As an example the
list of different geometries includes rectangular channels
obtained by hot embossing in polymer wafers, semi-circular
channels in isotropically etched surfaces, triangular channels in
KOH-etched silicon crystals, Gaussian-shaped channels in
laser-ablated polymer films, and elliptic channels in stretched
soft polymer PDMS devices~\cite{Geschke:04a}.

In this paper we introduce our recent
attempts~\cite{Mortensen:05b,Mortensen:05e} in giving a general
account for the mass and charge transport coefficients for an
electrolyte in a micro or nanochannel of arbitrary cross sectional
shape. To further motivate this work we emphasize that the flow of
electrolytes in the presence of a zeta potential is a scenario of
key importance to lab-on-a-chip applications involving biological
liquids/samples in both
microfluidic~\cite{Schasfoort:1999,Takamura:03,Reichmuth:03} and
nanofluidic
channels~\cite{Daiguji:2004,Stein:2004,Vanderheyden:2005,Brask:05a,Yao:03a,Yao:03b,Plecis:2005,Schoch:2005,Schoch:2005a,Jarlgaard:06}.

\begin{figure}[b]
\begin{center}
\epsfig{file=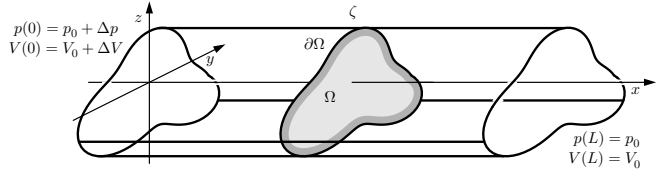, width=\columnwidth, clip}
\end{center}
\caption{A translation invariant channel of arbitrary cross
section $\Omega$ of area $\cal A$ containing an electrolyte driven
by a pressure gradient $-\Delta p/L$ and by electro-osmosis
through the potential gradient $-\Delta V/L$. The channel wall
$\partial \Omega$ has the electrical potential $\zeta$, which
induces a thin, charged Debye layer (dark gray) that surrounds the
charge neutral bulk (light gray). \label{fig1} }
\end{figure}

\section{Linear-response transport coefficients}

The general steady-state flow problem is illustrated in
Fig.~\ref{fig1} where pressure gradients and electro-osmosis (EO)
are playing in concert~\cite{ajdari:04a}. We consider a long,
straight channel of length $L$ having a constant cross section
$\Omega$ of area $\cal A$ and boundary $\partial\Omega$ of length
${\cal P}$. For many purposes it is natural to introduce a single
characteristic length scale

\begin{equation}
{\cal R}= \frac{2{\cal A}}{\cal P}
\end{equation}
which in the context of hydrodynamics is recognized as half the
hydraulic diameter. Indeed, for a circle of radius $R$ this gives
${\cal R}=R$.

The channel contains an incompressible electrolyte, which we for
simplicity assume to be binary and symmetric, i.e., containing
ions of charge $+Ze$ and $-Ze$ and equal diffusivities $D$. The
electrolyte has viscosity $\eta$, permittivity $\epsilon$, Debye
screening length $\lambda_D$, and bulk conductivity
$\sigma_\mathrm{\!o}=\epsilon D/\lambda_D^2$ and at the boundary
$\partial\Omega$ it has a zeta potential $\zeta$. The laminar,
steady-state transport of mass and charge is driven by a linear
pressure drop $\Delta p$ and a linear voltage drop $\Delta V$.
With these definitions flow will be in the positive $x$ direction.
In the linear-response regime the corresponding volume flow rate
$Q$ and charge current $I$ are related to the driving fields by
 \begin{equation}\label{eq:G}
 \left(\begin{array}{cc} Q \\ I
 \end{array}\right) =G\left(\begin{array}{cc}
 \Delta p \\ \Delta V
 \end{array}\right),\quad
 G=\left(\begin{array}{cc}G_{11}&G_{12}\\G_{21}&G_{22} \end{array}\right),
 \end{equation}
where, according to Onsager relations~\cite{Brunet:2004}, $G$ is a
symmetric, $G_{12}=G_{21}$, two-by-two conductance matrix. In the
following we introduce the characteristic conductance elements

 \begin{equation}\label{eq:G*}
 G^*=\left(\begin{array}{cc}G_{\rm hyd}^*&G_{\rm eo}^*\\G_{\rm eo}^*&G_{\rm mig}^* \end{array}\right)=\frac{\cal A}{L}\left(\begin{array}{cc}\frac{{\cal R}^2}{8\eta}\:&- \frac{\epsilon\zeta}{\eta}\\- \frac{\epsilon\zeta}{\eta}&\sigma_0 \end{array}\right),
 \end{equation}
which is the well-known result for a channel of circular cross
section of radius $R={\cal R}\gg \lambda_D$.

\section{Summary of results}

In the following we summarize our results for the transport
coefficients accompanied by more heuristic arguments before we in
the subsequent sections offer more detailed calculations. The
upper diagonal element is the hydraulic conductance or inverse
hydraulic resistance which to a good approximation is given by
 \begin{equation}\label{eq:G11_alpha}
 G_{11} \approx G_{\rm hyd}^*.
 \end{equation}
While there is no intrinsic length scale influencing $G_{11}$, the
other elements of $G$ depend on the Debye screening length
$\lambda_D$. This length can be comparable to and even exceed the
transverse dimensions in
nano-channels~\cite{Daiguji:2004,Stein:2004,Vanderheyden:2005}, in
which case the off-diagonal elements may depend strongly on the
actual cross-sectional geometry. However, for thin Debye layers
with a vanishing overlap the matrix elements $G_{12}$, $G_{21}$,
and $G_{22}$ are independent of the details of the geometry. For a
free electro-osmotic flow, a constant velocity field
$v^{{}}_\mathrm{eo} = (\epsilon \zeta/\eta)\Delta V/L$ is
established throughout the channel, except for in the thin Debye
layer of vanishing width. Hence $Q = v^{{}}_\mathrm{eo}{\cal A}$
and
\begin{subequations}
 \begin{equation}
 G_{12}= G_{21} =G_{\rm eo}^*,
 \quad \lambda_D \ll {\cal R}.
 \label{eq:G12thin}
 \end{equation}
From Ohm's law $I = (\sigma_\mathrm{\!o}{\cal A}/L)\Delta V$ it
follows that
 \begin{equation}
 G_{22} = G_{\rm mig}^*,
 \quad \lambda_D \ll {\cal R}.
 \label{eq:G22thin}
 \end{equation}
 \end{subequations}
For strongly overlapping Debye layers we shall see that in general
\begin{subequations}
 \begin{align}
 G_{12}= G_{21} &\approx \frac{{\cal R}^2}{8\lambda_D^2}\:G_{\rm eo}^*, \quad
\lambda_D\gg {\cal R},
 \label{eq:G12strong}\\
 G_{22} &=G_{\rm mig}^* + {\cal
 O}({\cal R}^2/\lambda_D^2)
 ,\quad \lambda_D\gg {\cal R}.
 \label{eq:G22strong}
 \end{align}
 \end{subequations}
We emphasize that the above results are generally valid for
symmetric electrolytes as well as for asymmetric electrolytes. We
also note that the expressions agree fully with the corresponding
limits for a circular cross section and the infinite parallel
plate system, were explicit solutions exist in terms of Bessel
functions~\cite{Rice:65,Probstein:94a} and cosine hyperbolic
functions~\cite{Probstein:94a}, respectively.
From the corresponding resistance matrix $R=G^{-1}$ we get the
hydraulic resistance
\begin{subequations}
 \begin{equation}
 R_{11}\approx \frac{1}{1-\beta}\frac{1}{G_{\rm hyd}^*},
 \end{equation}
where $\beta\equiv G_{12}G_{21}/(G_{11}G_{22})$ is the Debye-layer
correction factor to the hydraulic resistance. In the two limits
we have
 \begin{equation}
 \beta\approx\frac{8\epsilon^2\zeta^2}{\eta\sigma_\mathrm{\!o} {\cal R}^2 }\times
 \left\{\begin{array}{ccc}1&,&\displaystyle\lambda_D\ll
 {\cal R}\\
 \\ \displaystyle
 \Big( \frac{{\cal R}^2}{8\lambda_D^2}\Big)^2 &,&\displaystyle\lambda_D\gg
 {\cal R}\end{array}\right.
 \end{equation}
\end{subequations}
For $\zeta$ going to zero $\beta$ vanishes and we recover the
usual result for the hydraulic resistance.

\section{Governing equations}

For the system illustrated in Fig.~\ref{fig1}, an external
pressure gradient $\nablabf p = -(\Delta p/L)\ex$ and an external
electrical field $\EEE=E \ex=(\Delta V/L)\ex$ is applied. There is
full translation invariance along the $x$ axis, from which it
follows that the velocity field is of the form
$\vvv(\rrr)=v(\rrrp)\ex$ where $\rrrp = y\ey+z\ez$. For the
equilibrium potential and the corresponding charge density we have
$\phi_\mathrm{eq}(\rrr)=\phi_\mathrm{eq}(\rrrp)$ and
$\rho_\mathrm{eq}^e(\rrr)=\rho_\mathrm{eq}^e(\rrrp)$,
respectively. We will use the Dirac \emph{bra-ket}
notation~\cite{Dirac:81,Merzbacher:70} which is mainly appreciated
by researchers with a background in quantum physics, but as we
shall see it allows for a very compact, and in our mind elegant,
description of the present purely classical transport problem. In
the following functions $f(\rrrp)$ in the domain $\Omega$ are
written as $\big|f\big>$ with inner products defined by the
cross-section integral
 \begin{equation}\label{eq:inner}
 \big< f \big|g\big>\equiv \int_\Omega d\rrrp\, f(\rrrp)g(\rrrp).
 \end{equation}
From the Navier--Stokes equation it follows that the velocity of
the laminar flow is governed by the following force
balance~\cite{Batchelor:67,Landau:87a}
 \begin{equation}\label{eq:NS}
 0=\frac{\Delta p}{L}\big|1\big>+\eta \nabla_{\perp}^2
 \big|v\big>+\frac{\Delta V}{L} \big|\rho_\mathrm{eq}^e\big>,
\end{equation}
where $\nabla_{\perp}^2 = \partial^2_y + \partial^2_z$ is the 2D
Laplacian and $\big|1\big>$ corresponds to the unit function, i.e.
$g(\rrrp)=1$. The first term is the force-density from the
pressure gradient, the second term is viscous force-density, and
the third term is force-density transferred to the liquid from the
action of the electrical field on the electrolyte ions. The
equilibrium potential $\big|\phi_\mathrm{eq}\big>$ and the charge
density $\big|\rho_\mathrm{eq}^e\big>$ are related by the Poisson
equation
 \begin{equation}\label{eq:P}
 \nabla_{\perp}^2 \big|\phi_\mathrm{eq}\big> =-
 \frac{1}{\epsilon}\big|\rho_\mathrm{eq}^e \big>.
 \end{equation}
The velocity $\big|v\big>$ is subject to a no-slip boundary
condition on $\partial\Omega$ while the equilibrium potential
$\big|\phi_\mathrm{eq}\big>$ equals the zeta potential $\zeta$ on
$\partial\Omega$. Obviously, we also need a statistical model for
the electrolyte, and in the subsequent sections we will use the
Boltzmann model where the equilibrium potential
$\big|\phi_\mathrm{eq}\big>$ is governed by the Poisson--Boltzmann
equation. However, before turning to a specific model we will
first derive general results which are independent of the
description of the electrolyte.

We first note that because Eq.~(\ref{eq:NS}) is linear we can
decompose the velocity as $\big|v\big> =
\big|v_p\big>+\big|v_\mathrm{eo}\big>$, where $\big|v_p\big>$ is
the Hagen--Poiseuille pressure driven velocity governed by
 \begin{equation}
 0=\frac{\Delta p}{L}\big|1\big>+\eta \nabla_{\perp}^2
 \big|v_p\big>,
 \end{equation}
and $\big|v_\mathrm{eo}\big>$ is the electro-osmotic velocity
given by
 \begin{equation}\label{eq:veo}
 \big|v_\mathrm{eo}\big> = -\frac{\epsilon\Delta V}{\eta L}
 \big(\zeta\big|1\big>-\big|\phi_\mathrm{eq}\big>\big).
 \end{equation}
The latter result is obtained by substituting Eq.~(\ref{eq:P}) for
$\big|\rho_\mathrm{eq}^e\big>$ in Eq.~(\ref{eq:NS}). The upper
diagonal element in $G$ is given by
$G_{11}=\big<1\big|v_p\big>/\Delta p$ which may be parameterized
according to Eq.~(\ref{eq:G11_alpha}). The upper off-diagonal
element is given by $G_{12}=\big<1\big|v_\mathrm{eo}\big>/\Delta
V$ and combined with the Onsager relation we get
\begin{equation}\label{eq:G12_general}
G_{12}=G_{21}=-\frac{1}{L}\frac{\epsilon}{\eta}\big<1\big|\zeta-\phi_\mathrm{eq}\big>=-\frac{{\cal
A}}{L}\frac{\epsilon}{\eta}\big(\zeta-\bar\phi_\mathrm{eq}\big),
\end{equation}
where we have used that $\big< 1\big|1\big>={\cal A}$ and
introduced the average potential $\bar\phi_\mathrm{eq}=\big<
\phi_\mathrm{eq}\big|1\big>/\big<1\big|1\big>$.

There are two contributions to the lower diagonal element
$G_{22}$; one from migration,
$G_{22}^\mathrm{mig}=\big<1\big|\sigma\big>/L$, and one from
electro-osmotic convection of charge, $G_{22}^\mathrm{conv}=\big<
\rho_\mathrm{eq}^e\big|v_{\rm eo}\big>/\Delta V$, so that
\begin{equation}\label{eq:G22_general}
G_{22}=G_{22}^\mathrm{mig}+G_{22}^\mathrm{conv}=
\frac{1}{L}\big<1\big|\sigma\big>-\frac{\epsilon}{\eta L} \big<
\rho_\mathrm{eq}^e\big|\zeta-\phi_\mathrm{eq}\big>,
\end{equation}
where the electrical conductivity $\sigma(\rrrp)$ depends on the
particular model for the electrolyte. For thin non-overlapping
Debye layers we note that $\bar\phi_\mathrm{eq}\simeq 0$ so that
Eq.~(\ref{eq:G12_general}) reduces to Eq.~(\ref{eq:G12thin}) and,
similarly since the induced charge density is low,
Eq.~(\ref{eq:G22_general}) reduces to Eq.~(\ref{eq:G22thin}). For
strongly overlapping Debye layers the weak screening means that
$\phi_\mathrm{eq}$ approaches $\zeta$ so that the off-diagonal
elements $G_{12}=G_{21}$ and the $G_{22}^\mathrm{conv}$ part of
$G_{22}$ vanish entirely. In the following we consider a
particular model for the electrolyte and calculate the asymptotic
suppression as a function of the Debye screening length
$\lambda_D$.

\begin{table*}[t!]
\begin{tabular}{lcccccc}
& $\left(\kappa_1 {\cal R}\right)^2$ & ${\cal
A}_1^\mathrm{eff}/{\cal A}$& $\alpha$& $\gamma$
\\\hline
circle & $\gamma_1^2\simeq 5.78$$^{a,b}$ & $4/\gamma_1^2\simeq 0.69$$^{a,b}$  &  $4\pi$& 1$^c$\\
quarter-circle  & 5.08$^d$ & 0.65$^d$  & $29.97$$^d$ & 0.93$^d$ \\
half-circle     & 5.52$^d$ & 0.64$^d$ & $33.17$$^d$& 0.99$^d$ \\
ellipse(1:2)    & 6.00$^d$ & 0.67$^d$ & $10\pi$$^c$& 1.05$^d$\\
ellipse(1:3)    & 6.16$^d$ & 0.62$^d$  & $40\pi/3$$^c$& 1.11$^d$\\
ellipse(1:4)    & 6.28$^d$ & 0.58$^d$  & $17\pi$$^c$&
1.14$^d$\\\hline
triangle(1:1:1) & $4\pi^2/9\simeq 4.39$$^e$& $6/\pi^2\simeq 0.61$$^e$  & $20\sqrt{3}\:$$^c$ & $5/6\simeq 0.83$$^c$\\
triangle(1:1:$\sqrt{2}$) & $\frac{5\pi^2}
  {( 2 + \sqrt{2})^2}\simeq 4.23$$^a$ &$512/9\pi^4\simeq 0.58$$^a$
& $38.33$$^d$&0.82$^d$\\\hline
square(1:1) & $\pi^2/2\simeq 4.93$$^a$ & $64/\pi^4\simeq 0.66$$^a$  & $28.45$$^d$ &  0.89$^d$\\
rectangle(1:2) & $5\pi^2/9\simeq 5.48$$^a$& $64/\pi^4\simeq 0.66$$^a$  &  $34.98$$^d$& 0.97$^d$\\
rectangle(1:3) & $ 5\pi^2/8\simeq 6.17$$^a$ & $64/\pi^4\simeq 0.66$$^a$ &  $45.57$$^d$& 1.07$^d$\\
rectangle(1:4) & $17\pi^2/25\simeq 6.71$$^a$ &$64/\pi^4\simeq 0.66$$^a$ & $56.98$$^d$ &  1.14$^d$ \\
rectangle(1:$\infty$) & $ \sim\pi^2\simeq 9.87$$^a$
&$64/\pi^4\simeq 0.66$$^a$ & $\infty$ & $\sim 3/2$$^f$\\\hline
pentagon &  5.20$^d$ & 0.67$^d$& $26.77$$^d$ &0.92$^d$\\\hline
hexagon& 5.36$^d$ & 0.68$^d$ & $26.08$$^d$& 0.94$^d$\\\hline
\end{tabular}
\caption{Central dimensionless parameters for different
geometries. $^a$See e.g.~\cite{Morse:1953} for the eigenmodes and
eigenspectrum. $^b$Here, $\gamma_{1}\simeq 2.405$ is the first
root of the zeroth Bessel function of the first kind. $^c$See
e.g.~\cite{Mortensen:05b} and references therein. $^d$Data
obtained by finite-element simulations~\cite{comsol}. $^e$See
e.g.~\cite{Brack:1997} for the eigenmodes and eigenspectrum.
$^f$See e.g.~\cite{Batchelor:67} for a solution of the Poisson
equation.} \label{tab:1}
\end{table*}

\section{Debye--H\"uckel approximation}

Here we will limit ourselves to the Debye--H{\"u}ckel
approximation while more general results beyond that approximation
can be found in Ref.~\cite{Mortensen:05e}. In the
Debye--H{\"u}ckel approximation the equilibrium potential
$\big|\phi_\mathrm{eq}\big>$ is governed by the linearized
Poisson--Boltzmann equation~\cite{Squires:05a}
 \begin{equation}\label{eq:PB_DH}
\nabla_{\perp}^2\big|\phi_\mathrm{eq}\big> =
 \frac{1}{\lambda_D^2}\big|\phi_\mathrm{eq}\big>,
 \end{equation}
where $\lambda_D$ is the Debye screening length which for a
symmetric electrolyte is given by
\begin{equation}
\lambda_D= \sqrt{\frac{\epsilon k_B T}{2 (Ze)^2 c_\mathrm{o}}}
\end{equation}
with bulk concentration $c_\mathrm{o}$. The Debye--H{\"u}ckel
approximation is valid in the limit $Z\zeta e\ll k_BT$ where
thermal energy dominates over electrostatic energy. Since we
consider an open system connected to reservoirs at both ends of
the channel we are able to define a bulk equilibrium concentration
in the reservoirs even in the limit of strongly overlapping Debye
layers inside the channel. Thus, strongly overlapping Debye layers
do in this case not violate the underlying assumptions of the
Poisson--Boltzmann equation.

\begin{figure}[t!]
\begin{center}
\epsfig{file=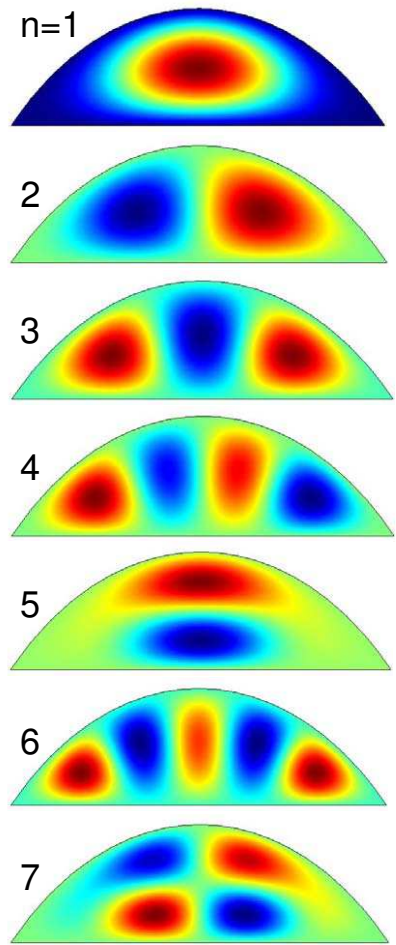, width=0.6\columnwidth, clip}
\end{center}
\caption{Examples of the first 7 eigenfunctions $\big|\psi_n\big>$
of Eq.~(\ref{eq:EigenvalDef}) with the eigenvalue $\kappa_n^2$
increasing with increasing $n$. For this particular case
$(\kappa_1 {\cal R})^2\simeq 5.05$ and ${\cal A}_1/{\cal A}\simeq
0.59$ while modes with $n=2$ and $n=4$ will in this case have
${\cal A}_n=0$ due to the symmetry.\label{fig2}}
\end{figure}

\subsection{Hilbert space formulation}
In order to solve Eqs.~(\ref{eq:NS}), (\ref{eq:P}), and
(\ref{eq:PB_DH}) we will take advantage of the Hilbert space
formulation~\cite{Morse:1953}, often employed in quantum
mechanics~\cite{Merzbacher:70}. The Hilbert space of real
functions on $\Omega$ is defined by the inner product in
Eq.~(\ref{eq:inner}) and a complete, countable set
$\big\{\big|\psi_n\big>\big\}$ of orthonormal basis functions,
i.e.,
 \begin{equation}
 \big<\psi_m\big|\psi_n\big>=\delta_{nm},
 \end{equation}
where $\delta_{nm}$ is the Kronecker delta. As our basis functions
we choose the eigenfunctions $\big\{\big|\psi_n\big>\big\}$ of the
Helmholtz equation with a zero Dirichlet boundary condition on
$\partial\Omega$,
 \begin{equation} \label{eq:EigenvalDef}
 -\nabla_{\perp}^2 \big|\psi_n\big> = \kappa_n^2 \big|\psi_n\big>,
 \quad n=1,2,3,\ldots.
 \end{equation}
The eigenstates of Eq.~(\ref{eq:EigenvalDef}) are well-known from
a variety of different physical systems including membrane
dynamics, the acoustics of drums, the single-particle eigenstates
of 2D quantum dots, and quantized conductance of quantum wires.
Furthermore, with an appropriate re-scaling of the Laplacian by
$\cal R$ or ${\cal A}/{\cal P}$ the lowest eigenvalue has a modest
dependence on the geometry~\cite{Mortensen:05c,Mortensen:05d}.
Fig.~\ref{fig2} shows as an example the 7 lowest eigenstates
$\big|\psi_n\big>$ in a particular geometry. With this complete
basis any function in the Hilbert space can be written as a linear
combination of basis functions. In the following we write the
fields as
\begin{subequations}
 \begin{align}
 \big|v\big>&=\sum_{n=1}^\infty a_n \big|\psi_n\big>,\label{eq:exp_v} \\
 \big|\phi_\mathrm{eq}\big>&=\zeta\big|1\big>-\sum_{n=1}^\infty b_n
 \big|\psi_n\big>,\label{eq:exp_phi}\\
 \big|\rho_\mathrm{eq}^e\big>&= \sum_{n=1}^\infty c_n
 \big|\psi_n\big>.\label{eq:exp_rho}
 \end{align}
 \end{subequations}
The linear problem is now solved by straightforward bra-ket
manipulations from which we identify the coefficients as
\begin{subequations}
 \begin{align} \label{eq:an}
 a_n &=\left(\frac{\Delta p}{\eta L}\frac{1}{\kappa_n^2}
 -\frac{\epsilon\zeta\Delta V}{\eta L}
 \frac{1}{1+(\kappa_n\lambda_D)^2}\right)\big<\psi_n\big|1\big>,\\
 \label{eq:bn}
 b_n &= \zeta
 \frac{\big<\psi_n\big|1\big>}{1+(\kappa_n\lambda_D)^2},
\\
 \label{eq:cn}
 c_n &=-\epsilon\zeta\kappa_n^2
 \frac{\big<\psi_n\big|1\big>}{1+(\kappa_n\lambda_D)^2}.
 \end{align}
 \end{subequations}

\begin{figure*}[t!]
\begin{center}
\epsfig{file=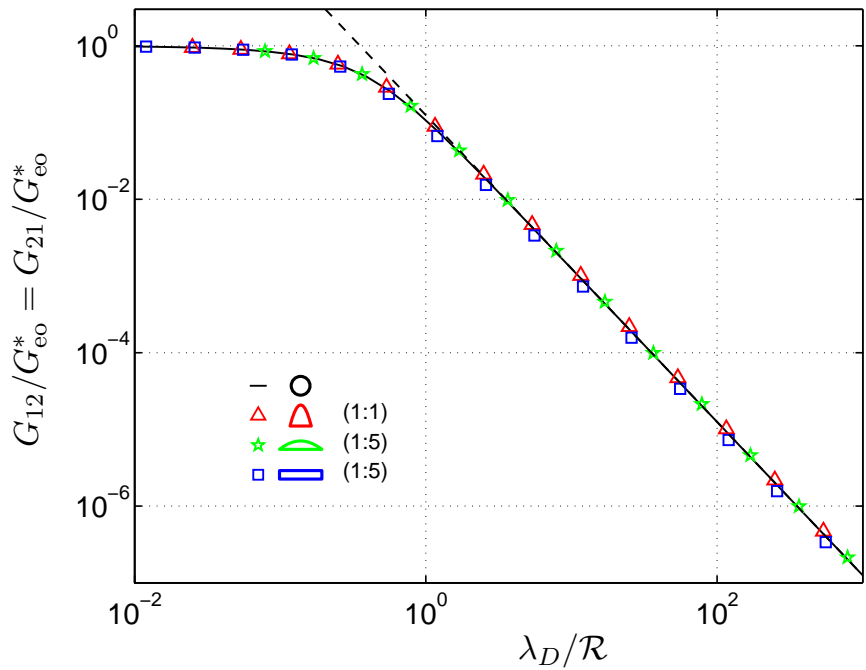, width=0.65\textwidth, clip}
\end{center}
\caption{Rescaled off-diagonal transport coefficients versus
rescaled Debye-layer thickness in the Debye--H\"uckel limit. The
solid line is the exact result for a circle,
Eq.~(\ref{eq:G12circle}), and the dashed line shows
Eq.~(\ref{eq:G12strong}). The data points are finite-element
simulations for different cross sections, see inset.\label{fig3}}
\end{figure*}

\subsection{Transport equations}

The flow rate and the electrical current are conveniently written
as
\begin{subequations}
 \begin{align}
 Q &=\big<1\big|v\big>,\label{eq:Q}\\
 I &= \big<\rho_\mathrm{eq}^e\big|v\big>+\sigma_\mathrm{\!o} E
 \big<1\big|1\big>,\label{eq:I}
 \end{align}
 \end{subequations}
where the second relation is the linearized Nernst--Planck
equation with the first term being the convection/streaming
current while the second is the ohmic current.

\subsection{Transport coefficients}

Substituting Eqs.~(\ref{eq:exp_v}) and (\ref{eq:exp_rho}) into
these expressions we identify the transport coefficients as

\begin{subequations}
 \begin{align}\label{eq:Ggeneral}
 G_{11}&= G_{\rm hyd}^* \sum_{n=1}^\infty \frac{8}{\left(\kappa_n {\cal R}\right)^2 }
   \frac{{\cal A}_n}{\cal A},\\
 G_{12} &=G_{\rm eo}^*
   \sum_{n=1}^\infty \frac{1}{1+(\kappa_n\lambda_D)^2}
   \frac{{\cal A}_n}{\cal A},\label{eq:G12general} \\
 G_{21} &=G_{\rm eo}^*
   \sum_{n=1}^\infty \frac{1}{1+(\kappa_n\lambda_D)^2}
   \frac{{\cal A}_n}{\cal A},\\
 G_{22}&=G_{\rm mig}^*+
   \frac{(\epsilon\zeta)^2}{\eta\lambda_D^2}
   \frac{{\cal A}}{L}\sum_{n=1}^\infty
   \frac{(\kappa_n\lambda_D)^2}{\big[1+(\kappa_n\lambda_D)^2\big]^2}
   \frac{{\cal A}_n}{\cal A}, \label{eq:Ggeneral:4}
 \end{align}
\end{subequations}

where
 \begin{equation}
 {\cal A}_n\equiv
 \frac{\big|\big<1\big|\psi_n\big>\big|^2}{\big<\psi_n\big|\psi_n\big>}
 =\big|\big<1\big|\psi_n\big>\big|^2
 \end{equation}
is the effective area of the eigenfunction $\big|\psi_n\big>$. The
ratio ${\cal A}_n/{\cal A}$ is consequently a measure of the
relative area occupied by $\big|\psi_n\big>$ satisfying the
sum-rule $\sum_{n=1}^\infty {\cal A}_n = {\cal A}$. We note that
as expected $G$ obeys the Onsager relation $G_{12}=G_{21}$.
Furthermore, using that
 \begin{equation}
 \frac{(\kappa_n\lambda_D)^2}{\big[1+(\kappa_n\lambda_D)^2\big]^2}
 = -\frac{\lambda_D}{2}
 \frac{\partial}{\partial\lambda_D}\frac{1}{1+(\kappa_n\lambda_D)^2},
 \end{equation}
we get the following bound between the off-diagonal elements
$G_{12}=G_{21}$ and the lower diagonal element $G_{22}$,
 \begin{equation}\label{eq:G22_G12}
 G_{22}=G_{\rm mig}^*+ \frac{\epsilon\zeta}{2\lambda_D}
 \frac{\partial G_{12}}{\partial\lambda_D}.
 \end{equation}

\subsection{Asymptotics and limiting cases}

\subsubsection{The geometrical correction factor}
In analogy with Ref.~\cite{Mortensen:05b} we define a geometrical
correction factor $\gamma\equiv G_{\rm hyd}^*/G_{11}$ which from
Eq.~(\ref{eq:Ggeneral}) becomes
\begin{equation}\label{eq:alpha}
\gamma \equiv\left(\sum_{n=1}^\infty \frac{8}{\left(\kappa_n {\cal
R}\right)^2} \frac{{\cal A}_n}{\cal A}\right)^{-1}\approx
\frac{\left(\kappa_1 {\cal R}\right)^2}{8} \frac{\cal A}{{\cal
A}_1}.
\end{equation}
Its relation to the dimensionless parameter $\alpha$ in
Ref.~\cite{Mortensen:05b} is $\gamma= \alpha/(2{\cal C})$ where
${\cal C}={\cal P}^2/{\cal A}$ is the compactness. As we shall see
$\gamma$ is of the order unity and only weakly dependent on the
geometry so that Eq.~(\ref{eq:G11_alpha}) is a good approximation
for the general result in Eq.~(\ref{eq:Ggeneral}).

\subsubsection{Non-overlapping, thin Debye layers}

For the off-diagonal elements of $G$ we use that
$[1+(\kappa_n\lambda_D)^2]^{-1}= 1 +{\cal
O}[(\kappa_n\lambda_D)^2]$. In Section~\ref{sec:numerics} we
numerically justify that the smallest dimensionless eigenvalue
$\kappa_1^2$ is of the order $1/{\cal R}^2$, so we may approximate
the sum by a factor of unity, see Table~\ref{tab:1}. If we
furthermore use that $\gamma\approx 1$ we arrive at
Eq.~(\ref{eq:G12thin}) for $\lambda_D\ll {\cal R}$. These results
for the off-diagonal elements are fully equivalent to the
Helmholtz--Smoluchowski result~\cite{Probstein:94a}. For $G_{22}$
we use that $(\kappa_n\lambda_D)^2[1+(\kappa_n\lambda_D)^2]^{-2}=
{\cal O}[(\kappa_n\lambda_D)^2]$, thus we may neglect the second
term, whereby we arrive at Eq.~(\ref{eq:G22thin}).

\subsubsection{Strongly overlapping Debye layers}
\label{sec:DH_strongoverlap}

In the case of $\kappa_1\lambda_D\gg 1$ we may use the result
$[1+(\kappa_n\lambda_D)^2]^{-1}=(\kappa_n\lambda_D)^{-2}+{\cal O}[
( \kappa_n \lambda_D)^{-4}]$ which together with $\gamma\approx 1$
gives Eq.~(\ref{eq:G12strong}) for strongly overlapping Debye
layers. For $G_{22}$ we use Eq.~(\ref{eq:G22_G12}) and arrive at
the result in Eq.~(\ref{eq:G22strong}).

\subsubsection{The circular case}

For a circular cross-section it can be shown
that~\cite{Probstein:94a}
 \begin{equation}\label{eq:G12circle}
 G_{12}^\mathrm{circ}
 =G_{21}^\mathrm{circ}=G_{\rm eo}^*\:
 \frac{I_2\!\big({\cal R}/\lambda_D\big)}{I_0\!\big({\cal R}/\lambda_D\big)},
 \end{equation}
where $I_n$ is the $n$th modified Bessel function of the first
kind, and were we have explicitly introduced the variable ${\cal
R}$ to emphasize the asymptotic dependence in
Eq.~(\ref{eq:G12strong}) for strongly overlapping Debye layers. We
note that we recover the limits in Eqs.~(\ref{eq:G12thin}) and
(\ref{eq:G12strong}) for $\lambda_D\ll {\cal R}$ and $\lambda_D\gg
{\cal R}$, respectively.

\section{Numerical results}
\label{sec:numerics}

\subsection{The Helmholtz basis}

Only few geometries allow analytical solutions of both the
Helmholtz equation and the Poisson equation. The circle is of
course among the most well-known solutions and the equilateral
triangle is another example. However, in general the equations
have to be solved numerically, and for this purpose we have used
the commercially available finite-element software Comsol
Multiphysics~\cite{comsol}. Fig.~\ref{fig2} shows the results of
finite-element simulations for a particular geometry. The first
eigenstate of the Helmholtz equation is in general non-degenerate
and numbers for a selection of geometries are tabulated in Table
1. Note how the different numbers converge when going through the
regular polygons starting from the equilateral triangle through
the square, the regular pentagon, and the regular hexagon to the
circle. In general, $\kappa_1^2$ is of the order $1/{\cal R}^2$,
and for relevant high-order modes (those with a nonzero ${\cal
A}_n$) the eigenvalue is typically much larger. Similarly, for the
effective area we find that ${\cal A}_1/{\cal A}\leq
4/\gamma_1^2\simeq 0.69$ and consequently we have ${\cal
A}_n/{\cal A}< 1- 4/\gamma_1^2\simeq 0.31$ for $n\geq 2$.

The transport coefficients in Eqs.~(\ref{eq:Ggeneral})
to~(\ref{eq:Ggeneral:4}) are thus strongly influenced by the first
eigenmode which may be used for approximations and estimates of
the transport coefficients. As an example the column for $\gamma$
is well approximated by only including the first eigenvalue in the
summation in Eq.~(\ref{eq:alpha}). In fact, the approximation
$\gamma\approx 1$ is indeed reasonable.

\subsection{Transport coefficients}

Our analytical results predict that when going to either of the
limits of thin non-overlapping or strongly overlapping Debye
layers, the transport coefficients to a good approximation only
depend on the channel geometry through the hydraulic radius ${\cal
R}$. Therefore, when plotted against the rescaled Debye length
$\lambda_D/{\cal R}$, all our results should collapse on the same
asymptotes in the two limits.

In Fig.~\ref{fig3} we show the results for the off-diagonal
coefficients obtained from finite-element simulations in the
Debye--H\"uckel limit for three different channel cross sections,
namely two parabola shaped channels of aspect ratio 1:1 and 1:5,
respectively, and a rectangular channel of aspect ratio 1:5. In
all cases we find excellent agreement between the numerics and the
asymptotic expressions. For the comparison we have also included
exact results, Eq.~(\ref{eq:G12circle}), for the circular cross
section as well as results based on only the first eigenvalue in
Eq.~(\ref{eq:G12general}). Even though Eq.~(\ref{eq:G12circle}) is
derived for a circular geometry we find that it also accounts
remarkably well for even highly non-circular geometries in the
intermediate regime of weakly overlapping Debye layers.

\section{Conclusion}

We have analyzed the flow of incompressible electrolytes in long,
straight channels driven by pressure and electro-osmosis. By using
a powerful Hilbert space eigenfunction expansion we have been able
to address the general problem of an arbitrary cross section and
obtained general results for the hydraulic and electrical
transport coefficients. Results for strongly overlapping and thin,
non-overlapping Debye layers are particular simple, and from these
analytical results we have calculated the corrections to the
hydraulic resistance due to electro-hydrodynamic interactions.
These analytical results reveal that the geometry dependence only
appears through the hydraulic radius ${\cal R}$ and the correction
factor $\gamma$, as the expressions only depend on the rescaled
Debye length $\lambda_D/{\cal R}$ and $\gamma\approx 1$. Our
numerical analysis based on finite-element simulations indicates
that these conclusions are generally valid also for intermediate
values of $\lambda_D$. The present results constitute an important
step toward circuit analysis~\cite{Brask:03a,ajdari:04a} of
complicated micro and nanofluidic networks incorporating
complicated cross-sectional channel geometries.

\section*{Acknowledgments}

We thank Henrik Flyvbjerg for stimulating discussions which led to
the present definition of the geometrical correction factor
$\gamma$. This work is supported by the \emph{Danish Technical
Research Council} (Grant Nos.~26-03-0073~and~26-03-0037) and by
the \emph{Danish Council for Strategic Research} through the
\emph{Strategic Program for Young Researchers} (Grant No.:
2117-05-0037).


\begin{thebibliography}{10}
\providecommand{\url}[1]{#1} \csname url@rmstyle\endcsname
\providecommand{\newblock}{\relax}
\providecommand{\bibinfo}[2]{#2}
\providecommand\BIBentrySTDinterwordspacing{\spaceskip=0pt\relax}
\providecommand\BIBentryALTinterwordstretchfactor{4}
\providecommand\BIBentryALTinterwordspacing{\spaceskip=\fontdimen2\font
plus \BIBentryALTinterwordstretchfactor\fontdimen3\font minus
  \fontdimen4\font\relax}
\providecommand\BIBforeignlanguage[2]{{%
\expandafter\ifx\csname l@#1\endcsname\relax
\typeout{** WARNING: IEEEtran.bst: No hyphenation pattern has been}%
\typeout{** loaded for the language `#1'. Using the pattern for}%
\typeout{** the default language instead.}%
\else \language=\csname l@#1\endcsname \fi #2}}

\bibitem{Laser:04}
D.~J. Laser and J.~G. Santiago, ``A review of micropumps,''
  \emph{J.~Micromech.~Microeng.}, vol.~14, no.~6, pp. R35 -- R64, 2004.

\bibitem{Stone:04a}
H.~A. Stone, A.~D. Stroock, and A.~Ajdari, ``Engineering flows in
small
  devices: Microfluidics toward a lab-on-a-chip,''
  \emph{Annu.~Rev.~Fluid~Mech.}, vol.~36, pp. 381 -- 411, 2004.

\bibitem{Squires:05a}
T.~M. Squires and S.~R. Quake, ``Microfluidics: Fluid physics at
the nanoliter
  scale,'' \emph{Rev. Mod. Phys.}, vol.~77, pp. 977 -- 1026, 2005.

\bibitem{Geschke:04a}
O.~Geschke, H.~Klank, and P.~Telleman, Eds., \emph{Microsystem
Engineering of
  Lab-on-a-Chip Devices}.\hskip 1em plus 0.5em minus 0.4em\relax Weinheim:
  Wiley-VCH Verlag, 2004.

\bibitem{Mortensen:05b}
N.~A. Mortensen, F.~Okkels, and H.~Bruus, ``Reexamination of
  \uppercase{H}agen--\uppercase{P}oiseuille flow: Shape dependence of the
  hydraulic resistance in microchannels,'' \emph{Phys.~Rev.~E}, vol.~71, p.
  057301, 2005.

\bibitem{Mortensen:05e}
N.~A. Mortensen, L.~H. Olesen, and H.~Bruus, ``Transport
coefficients for
  electrolytes in arbitrarily shaped nano and micro-fluidic channels,''
  \emph{New J. Phys.}, vol.~8, p.~37, 2006.

\bibitem{Schasfoort:1999}
R.~B.~M. Schasfoort, S.~Schlautmann, L.~Hendrikse, and A.~{van den
Berg},
  ``Field-effect flow control for microfabricated fluidic networks,''
  \emph{Science}, vol. 286, no. 5441, pp. 942 -- 945, 1999.

\bibitem{Takamura:03}
Y.~Takamura, H.~Onoda, H.~Inokuchi, S.~Adachi, A.~Oki, and
Y.~Horiike,
  ``Low-voltage electroosmosis pump for stand-alone microfluidics devices,''
  \emph{Electrophoresis}, vol.~24, no. 1-2, pp. 185 -- 192, 2003.

\bibitem{Reichmuth:03}
D.~S. Reichmuth, G.~S. Chirica, and B.~J. Kirby, ``Increasing the
performance
  of high-pressure, high-efficiency electrokinetic micropumps using
  zwitterionic solute additives,'' \emph{Sens. Actuator B-Chem.}, vol.~92, no.
  1-2, pp. 37 -- 43, 2003.

\bibitem{Daiguji:2004}
H.~Daiguji, P.~D. Yang, A.~J. Szeri, and A.~Majumdar,
``Electrochemomechanical
  energy conversion in nanofluidic channels,'' \emph{Nano Lett.}, vol.~4,
  no.~12, pp. 2315 -- 2321, 2004.

\bibitem{Stein:2004}
D.~Stein, M.~Kruithof, and C.~Dekker, ``Surface-charge-governed
ion transport
  in nanofluidic channels,'' \emph{Phys.~Rev.~Lett.}, vol.~93, no.~3, p.
  035901, 2004.

\bibitem{Vanderheyden:2005}
F.~H.~J. {van der Heyden}, D.~Stein, and C.~Dekker, ``Streaming
currents in a
  single nanofluidic channel,'' \emph{Phys.~Rev.~Lett.}, vol.~95, no.~11, p.
  116104, 2005.

\bibitem{Brask:05a}
A.~Brask, J.~P. Kutter, and H.~Bruus, ``Long-term stable
electroosmotic pump
  with ion exchange membranes,'' \emph{Lab Chip}, vol.~5, no.~7, pp. 730 --
  738, 2005.

\bibitem{Yao:03a}
S.~H. Yao and J.~G. Santiago, ``Porous glass electroosmotic pumps:
theory,''
  \emph{J. Colloid Interface Sci.}, vol. 268, no.~1, pp. 133 -- 142, 2003.

\bibitem{Yao:03b}
S.~H. Yao, D.~E. Hertzog, S.~L. Zeng, J.~C. Mikkelsen, and J.~G.
Santiago,
  ``Porous glass electroosmotic pumps: design and experiments,'' \emph{J.
  Colloid Interface Sci.}, vol. 268, no.~1, pp. 143 -- 153, 2003.

\bibitem{Plecis:2005}
A.~Plecis, R.~B. Schoch, and P.~Renaud, ``Ionic transport
phenomena in
  nanofluidics: Experimental and theoretical study of the exclusion-enrichment
  effect on a chip,'' \emph{Nano Lett.}, vol.~5, no.~6, pp. 1147 -- 1155, 2005.

\bibitem{Schoch:2005}
R.~B. Schoch, H.~{van Lintel}, and P.~Renaud, ``Effect of the
surface charge on
  ion transport through nanoslits,'' \emph{Phys. Fluids}, vol.~17, no.~10, p.
  100604, 2005.

\bibitem{Schoch:2005a}
R.~B. Schoch and P.~Renaud, ``Ion transport through nanoslits
dominated by the
  effective surface charge,'' \emph{Appl. Phys. Lett.}, vol.~86, no.~25, p.
  253111, 2005.

\bibitem{Jarlgaard:06}
S.~E. Jarlgaard, M.~B.~L. Mikkelsen, P.~Skafte-Pedersen, H.~Bruus,
and
  A.~Kristensen, ``Capillary filling speed in silicon dioxide nano-channels,''
  in \emph{Proc. NSTI-Nanotech 2006}, vol.~2, 2006, pp. 521 -- 523.

\bibitem{ajdari:04a}
A.~Ajdari, ``Steady flows in networks of microfluidic channels:
building on the
  analogy with electrical circuits,'' \emph{C.~R.~Physique}, vol.~5, pp. 539 --
  546, 2004.

\bibitem{Brunet:2004}
E.~Brunet and A.~Ajdari, ``Generalized onsager relations for
electrokinetic
  effects in anisotropic and heterogeneous geometries,'' \emph{Phys.~Rev.~E},
  vol.~69, no.~1, p. 016306, 2004.

\bibitem{Rice:65}
C.~L. Rice and R.~Whitehead, ``Electrokinetic flow in a narrow
cylindrical
  capillary,'' \emph{J. Phys. Chem.}, vol.~69, no.~11, pp. 4017 -- 4024, 1965.

\bibitem{Probstein:94a}
R.~F. Probstein, \emph{PhysicoChemical Hydrodynamics, an
introduction}.\hskip
  1em plus 0.5em minus 0.4em\relax New-York: John Wiley and Sons, 1994.

\bibitem{Dirac:81}
P.~A.~M. Dirac, \emph{The Principles of Quantum Mechanics},
4th~ed.\hskip 1em
  plus 0.5em minus 0.4em\relax Oxford: Oxford University Press, 1981.

\bibitem{Merzbacher:70}
E.~Merzbacher, \emph{Quantum Mechanics}.\hskip 1em plus 0.5em
minus 0.4em\relax
  New York: Wiley \& Sons, 1970.

\bibitem{Batchelor:67}
G.~K. Batchelor, \emph{An Introduction to Fluid Dynamics}.\hskip
1em plus 0.5em
  minus 0.4em\relax Cambridge: Cambridge University Press, 1967.

\bibitem{Landau:87a}
L.~D. Landau and E.~M. Lifshitz, \emph{Fluid Mechanics}, 2nd~ed.,
ser. Landau
  and Lifshitz, Course of Theoretical Physics.\hskip 1em plus 0.5em minus
  0.4em\relax Oxford: Butterworth--Heinemann, 1987, vol.~6.

\bibitem{Morse:1953}
P.~M. Morse and H.~Feshbach, \emph{Methods of Theoretical
Physics}.\hskip 1em
  plus 0.5em minus 0.4em\relax New York: McGraw--Hill, 1953.

\bibitem{comsol}
Comsol support and Femlab documentation, www.comsol.com.

\bibitem{Brack:1997}
M.~Brack and R.~K. Bhaduri, \emph{Semiclassical Physics}.\hskip
1em plus 0.5em
  minus 0.4em\relax New York: Addison Wesley, 1997.

\bibitem{Mortensen:05c}
N.~A. Mortensen, F.~Okkels, and H.~Bruus, ``Universality in
edge-source
  diffusion dynamics,'' \emph{Phys. Rev. E}, vol.~73, p. 012101, 2006.

\bibitem{Mortensen:05d}
N.~A. Mortensen and H.~Bruus, ``Universal dynamics in the onset of
a
  hagen-poiseuille flow,'' \emph{Phys.~Rev.~E}, vol.~74, p. 017301, 2006.

\bibitem{Brask:03a}
A.~Brask, G.~Goranovi{\'{c}}, and H.~Bruus, ``Theoretical analysis
of the
  low-voltage cascade electroosmotic pump,'' \emph{Sens. Actuator B-Chem.},
  vol.~92, pp. 127--132, 2003.

\end{thebibliography}

\end{document}